\documentclass{sig-alternate}
\usepackage{balance}

\usepackage[
  pass,
]{geometry}

\usepackage{amsmath}
\usepackage{rotating}
\usepackage{tabularx}
\usepackage{booktabs}

\PassOptionsToPackage{hyphens}{url}\usepackage{hyperref}
\usepackage{cite} 
\usepackage{multirow}
\usepackage{array}
\usepackage[export]{adjustbox} 
\usepackage[framemethod=tikz]{mdframed}

\usepackage[export]{adjustbox}
\usepackage[hang]{subfigure}

\newmdenv[
  tikzsetting={draw=black,dashed,line width=0.5pt,dash pattern = on 4pt off 2pt},
  linecolor=white,
  backgroundcolor=white
]{dashedbox}
  
\newmdenv[
  tikzsetting={draw=black, line width=0.5pt},
  linecolor=black,
  backgroundcolor=white
]{normalbox}

\newcommand{\myCaption}[1]{\vspace{0.6\baselineskip}\noindent\textbf{#1}\vspace{0.4\baselineskip}}

\usepackage{float}
\usepackage{color}

\newcommand{\refRQ}[1]{\hfill {\tiny $\rightarrow RQ#1$}}

\newcommand{\refTab}[2]{{\scriptsize(#1,#2)}}

\newcommand{\comment}[1]{}

\begin{document}
%
\conferenceinfo{FSE'14}{, November 16--22, 2014, Hong Kong, China}
\CopyrightYear{2014} 
\crdata{978-1-4503-3056-5/14/11} 

\title{Sketches and Diagrams in Practice}
%
%
%
%
%

\numberofauthors{2} 
%
\author{
%
%
\alignauthor
Sebastian Baltes\\
       \affaddr{Computer Science}\\
 	\affaddr{University of Trier}\\
     	\affaddr{Trier, Germany}\\
    	\email{s.baltes@uni-trier.de}
\alignauthor
Stephan Diehl\\
  	\affaddr{Computer Science}\\
       \affaddr{University of Trier}\\
       	\affaddr{Trier, Germany}\\
       	\email{diehl@uni-trier.de}
}

\maketitle
\begin{abstract} 
Sketches and diagrams play an important role in the daily work of software developers.
In this paper, we investigate the use of sketches and diagrams in software engineering practice.
To this end, we used both quantitative and qualitative methods.
We present the results of an exploratory study in three companies and an online survey with 394 participants.
Our participants included software developers, software architects, project managers, consultants, as well as researchers.
They worked in different countries and on projects from a wide range of application areas.
Most questions in the survey were related to the last sketch or diagram that the participants had created.
Contrary to our expectations and previous work, the majority of sketches and diagrams contained at least some UML elements.
However, most of them were informal.
The most common purposes for creating sketches and diagrams were designing, explaining, and understanding, but analyzing requirements was also named often.
More than half of the sketches and diagrams were created on analog media like paper or whiteboards and have been revised after creation.
Most of them were used for more than a week and were archived.
We found that the majority of participants related their sketches to methods, classes, or packages, but not to source code artifacts with a lower level of abstraction.
\end{abstract}




\newpage

\section{Introduction}

Over the past years, studies have shown the importance of sketches and diagrams in software development ~\cite{Dekel07, Cherubini07, Walny11-1}.
Most of these visual artifacts do not follow formal conventions like the \emph{Unified Modeling Language} (UML), but have an informal, ad-hoc nature~\cite{Cherubini07, Petre13, Dekel07, Mangano14}.
Sketches and diagrams are important because they depict parts of the mental model developers build to understand a software project~\cite{LaToza06}.
They may contain different views, levels of abstraction, formal and informal notations, pictures, or generated parts~\cite{Hoek14, Dekel07, Cherubini07, Walny11-2}.
Developers create sketches and diagrams mainly to understand, to design, and to communicate~\cite{Cherubini07}.
Media for sketch creation include whiteboards, engineering notebooks, scrap papers, but also software tools like Photoshop and PowerPoint~\cite{Walny11-1, Cherubini07, Myers08}.
When designing, sketches relieve short-term memory, augment information processing, and are a source of creativity~\cite{Goldschmidt03, Suwa00, Tversky03}.

The goal of our research was to investigate the usage of sketches and diagrams in software engineering practice and their relation to the core elements of a software project, the source code artifacts.
Furthermore, we wanted to assess how helpful sketches are for understanding the related code.
We intended to find out if, how, and why sketches and diagrams are archived and are thereby available for future use.
Since software is created with and for a wide range of stakeholders~\cite{Taylor07} and sketches are often a means for communicating between these stakeholders, we were not only interested in sketches and diagrams created by software developers, but by all software practitioners, including testers, architects, project managers, as well as researchers and consultants.

Our study complements a number of existing studies on the use of sketches and diagrams in software development (see Section \ref{sec:related-work}),
which analyzed the above aspects only in parts and often focused on an academic environment~\cite{Walny11-1}, a single company~\cite{LaToza06, Cherubini07}, open source projects~\cite{Yatani09, Chung10}, or were limited to a small group of participants~\cite{Yatani09, Petre13}.
Based on our findings, we point at the need for tool support to better integrate sketches and diagrams into the software development process.
Throughout this paper, we summarize preliminary findings in boxes with dashed borders and our final results in boxes with solid borders.
For simplicity, we use the term \textit{sketch} in the following to denote both informal sketches as well as diagrams following a formal notation like UML.

\section{Research Design}

\begin{figure}[tb]
\centering
\includegraphics[width=0.7\columnwidth]{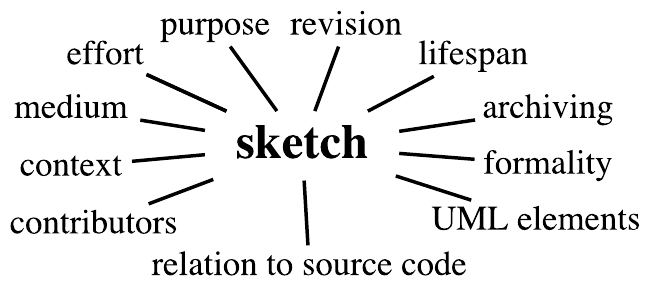}
\vspace{-\baselineskip}
\caption{Dimensions of a sketch or diagram in software development}
\label{fig:dimensions}
\vspace{-\baselineskip}
\end{figure}

Our research was carried out in two phases: 
First, we conducted an exploratory field study on the use of sketches and diagrams in three different software companies to determine a set of dimensions for characterizing the visual artifacts.
In total, we identified the 11 dimensions that are shown in Figure \ref{fig:dimensions}.
Some of them were derived from related work, others emerged during our research. 
In the second phase, we asked practitioners to describe their last sketch based on these dimensions in an online survey with 394 participants.

\subsection{Exploratory Research}

For our exploratory study, we collaborated with 
a company developing utility software (company A), 
a company developing software for real-time devices (company B), 
and a company developing software for the health care sector (company C).
Companies A and B are small to medium-sized enterprises, whereas company C is a large corporation.

\subsubsection{Field Study}

First, our interest focused on real-life sketches drawn by software practitioners. 
Previous studies showed that, in practice, sketches and diagrams are often rather informal~\cite{Dekel07, Cherubini07, Petre13}.
However, we had only seen few samples of sketches drawn by professional software developers.
Thus, we collected 47 sketches drawn by 13 different developers of companies A and B and interviewed them about certain properties of their sketches.

We prepared two questionnaires, one for each developer and one for each collected sketch.
Using the developer questionnaire, we captured demographic data like gender, age, and work experience.
Furthermore, we asked how often the participant normally uses sketches in his or her daily work.
For each sketch, we asked how many persons contributed to it, the purpose for creating the sketch, and its (estimated) lifespan.
Moreover, we requested a small description and asked for the relation of the sketch to source code.

The median age of the developers was 29 and their median work experience was 3 years.
We were surprised by the broad spectrum of sketches and diagrams, even in this limited setting.
The sketches ranged from simple to-do lists through visualizations of geometric problems to computer-generated UML diagrams.
However, the majority of sketches were informal, only two of them contained UML elements.
The most common purposes were understanding issues and designing new features. 
The median lifespan was rather short (2 to 3 days) and only a minority of sketches were kept permanently (8.5\%).
Developers related 79\% of their sketches to methods or classes.
In company A, the employees used sketches on a monthly basis, in company B on a weekly basis.
These results led to first preliminary assumptions on the dimensions \textit{formality}, \textit{UML elements}, \textit{purpose}, \textit{lifespan}, \textit{archiving}, and \textit{relation to source code}.

\begin{dashedbox}
Sketches are mostly informal and UML is rarely used.
Their main purposes are understanding and designing and the lifespan is in most cases only a few days.
Sketches are rarely archived and are mainly related to classes and methods.
\end{dashedbox}

\subsubsection{Interviews}

Since the questionnaires and collected sketches revealed differences between the cultures of the two companies regarding sketch usage, we wanted to investigate how sketching is integrated in the software development workflow of different companies.
Therefore, we semi-structurally interviewed one software developer and the chairman of company B.
As we could not interview employees of company A, we recruited two developers from company C to be interviewed.

The interviews revealed that the management of company B actively demanded sketching and sketch archiving, without forcing the developers to use a certain notation; in company C, the role of sketching differed between the teams.
One developer reported that in his team, informal whiteboard and paper sketches were used ``almost daily'' in different situations.
In contrast to that, the other developer from the same company, but a different team, noted that hand-drawn sketches were used ``surprisingly little''.

While the collected sketches were mostly created on paper, the interviews revealed the importance of whiteboards as sketching \textit{media}.
One participant reported that in his team, the whiteboard is normally used as soon as more than two persons are involved.
Otherwise, the preferred medium is paper, but computer programs like PowerPoint and Enterprise Architect are also applied.
As mentioned above, only few of the collected sketches were archived permanently.
However, all participants mentioned important sketches being archived, either by saving a digital picture or by redrawing them digitally.
The latter is an example of a transition from one medium to another, as described by Walny et al.~\cite{Walny11-1}.
We did not focus on these transitions in the field study, but decided to further investigate the reuse and \textit{revision} of sketches in our online survey.
In this context, we also wanted to assess how much \textit{effort} goes into the creation of sketches.
Furthermore, the interviews revealed the importance of the team \textit{context} and the \textit{contributors} that helped creating the sketch.

\begin{dashedbox}
Context and contributors influence the sketching practice in teams.
Paper and computers are used when sketching with one or two persons, otherwise whiteboards are the preferred medium.
\end{dashedbox}

\subsection{Research Questions}

While some of the sketch dimensions were already partly addressed in existing studies (e.g., medium~\cite{LaToza06, Cherubini07}, purpose~\cite{LaToza06, Cherubini07}, revision~\cite{Walny11-1}, and UML elements~\cite{Petre13}), others, like the relation to source code, have not been investigated yet.
With our survey, we wanted to reproduce findings from the other studies, but also intended to gain new insights into the use of sketches and diagram in software development practice.
We were especially interested in the reasons why or why not sketches are archived.
Moreover, we wanted to collect data on the actual lifespan of sketches in practice and their relation to source code, focussing on their value for documenting the related source code artifacts.
We dedicated an own dimension to UML, because it it often seen as the ``lingua franca'' of software engineering (see, e.g.,~\cite{Evans00}).
Since past studies and our preliminary research showed that UML is often used informally~\cite{Cherubini07, Petre13}, we wanted to assess how many UML elements are actually present on sketches created by software practitioners.
The following five research questions summarize what we wanted to learn with our online survey:

\begin{description}
\setlength{\itemsep}{0pt}
\item[RQ1] How frequently do developers create, use, and revise sketches and how much effort is put into these sketches?
\item[RQ2] What is the lifespan of sketches; to what extent and why are they archived?
\item[RQ3] How formal are sketches and to what extent do they contain UML elements?
\item[RQ4] What media are used for creating sketches, in what context and for what purpose are they created, and how many persons contribute to them?
\item[RQ5] To what source code artifacts are sketches related and could they help to understand these artifacts in the future?
\end{description}

\subsection{Survey Design}

To investigate our five research questions, we designed an online questionnaire consisting of 28 questions in total, 15 of which referred to the last sketch or diagram that the participant created for a professional software project.
Three of these questions were open-ended, the others were closed-ended.
One of our main goals while designing the questionnaire was to make it as concise as possible to increase the completion rate (see, e.g.,~\cite{Dillman93}) and to make it more likely that companies would forward our request for participation.  

To assess the 11 dimensions mentioned above, we asked the participants about the last sketch or diagram they created for a software project.
Furthermore, we asked if the sketch could be helpful in the future to understand the related source code artifact(s) and for demographic data about the respondent.
The target population for our study, i.e., the population of which we drew a sample, were basically all software practitioners in the world, meaning all software developers, testers, and architects, but also consultants, researcher, and everyone else involved in creation of software.

The questionnaire was online from August 28, 2013 until December 31, 2013, collecting 394 responses in total.
First, we recruited participants by a network of colleagues and contacts, asking them to motivate others to participate in our study.
In a second phase, we posted a call for participation in two social networks, various online communities and IRC channels.
Furthermore, we contacted several German software companies and asked them to forward a call for participation to their employees. 
In a third phase, the German IT news website \textit{heise developer} \cite{Heise14} published a short article on our survey, asking the readers to participate.
In the last recruitment phase, we contacted people working in the area of software engineering, asking them to advertise our survey on Twitter.
We also posted a call for participation in a large LinkedIn group with more than 44.000 members, focusing on software architecture.

All survey questions and the coding can be found in Table~\ref{tab:variables}.
Moreover, we made the questionnaire and the answers publicly available \cite{ST14}.
The variables in the table are either directly related to a certain research question or capture demographic data about the participants.
The names of most variables are based on the sketch dimensions.
However, we used multiple variables to capture the context in which the sketch was created (team size, application area, employment of agile methods and model-driven software engineering).

Beside these closed-ended questions, our questionnaire contained three open-ended questions:
Two of them were related to RQ2, asking for reasons why or why not the sketch has been archived.
At the end of our questionnaire, participants had the possibility to provide general remarks on their last sketch or their general usage of sketches and diagrams.

\begin{table*}[p!]
\renewcommand{\arraystretch}{1.1}
\scriptsize
\fontdimen2\font=2pt 
\centering
\vspace{-8pt}
\caption{Structure of online survey. Asterisks indicate level of measurement (no asterisk: nominal scale, one asterisk: ordinal scale, two asterisks: ratio scale).}
\begin{tabularx}{\textwidth}{| p{1.8cm} | >{\hsize=0.63\hsize}X | >{\hsize=1.37\hsize}X |} 
\hline
\textbf{Variable} & \textbf{Question} & \textbf{Values and Coding} \\
\hline
\textit{creation$^*$} \newline (CRE) \refRQ{1}
& When did you create your last sketch or diagram (that you created for your professional work and that is related to a software project)?
&   { \renewcommand{\arraystretch}{1}
       \hspace{-8pt}
       \begin{tabular}[t]{ll}
      0 = less than 10 minutes ago                                   	& 4 = several weeks ago (1-4 weeks)\\
      1 = several minutes ago (10-60 minutes)                  	& 5 = several months ago (1-12 month)\\
      2 = several hours ago (1-8 hours)                             	& 6 = more than one year ago\\
      3 = several work days ago (1-5 days)				& NA = I don't know\\
    \end{tabular}
   }\\
\hline
\textit{revision$^*$} \newline (REV) \refRQ{1}
& Has the sketch/diagram been revised after its initial creation?
&	{ \renewcommand{\arraystretch}{1}
         \hspace{-8pt}
       \begin{tabular}[t]{llll}
       0 = no  & 1 = yes, once  & 2 = yes, multiple times & NA = I don't know
	  \end{tabular}
    }\\
\hline
\textit{effort$^*$} \newline (EFF) \refRQ{1} 
& How much effective work time went into the creation and revision of the 
   sketch/diagram up to now? 
&	{ \renewcommand{\arraystretch}{1}
        \hspace{-8pt}
       \begin{tabular}[t]{ll}
       0 = less than 10 minutes & 3 = several work days (1-5 days)\\
	1 = several minutes (10-60 minutes) & 4 = more than 5 work days \\
	2 = several hours (1-8 hours) & NA = I don't know
	  \end{tabular}
    }\newline (If several persons were involved, add up the work times of all contributors.) \\	

\hline
\textit{contributors$^*$} \newline (CON) \refRQ{4}
& How many persons contributed to the sketch/diagram up to now (including yourself)? 
&	{ \renewcommand{\arraystretch}{1}
        \hspace{-8pt}
      \begin{tabular}[t]{lll}
      1 = 1 person  &  4 = 4 to 10 persons & 7 = more than 100 persons \\
	2 = 2 persons & 	5 = 11 to 50 persons & NA = I don't know\\ 
	3 = 3 persons & 	6 = 51 to 100 persons
  \end{tabular}
    }\\	
\hline
\textit{medium} \newline (MED) \refRQ{4}
& What medium did you use to create the sketch/diagram?
& paper / traditional whiteboard / interactive whiteboard / tablet or smartphone / computer / other \\
\hline
\textit{lifespan$^*$} \newline (LSP) \refRQ{2}
& Please try to estimate the lifespan of the sketch/diagram (how long did/will you use it?).
&	{ \renewcommand{\arraystretch}{1}
        \hspace{-8pt}
      \begin{tabular}[t]{ll}
       0 = lifespan ended immediately after creation & 4 = several work days (1-5 days)\\
	1 = less than 10 minutes & 5 = several weeks (1-4 weeks)\\
      2 = several minutes (10-60 minutes) & 6 = several months (1-12 months)\\
      3 = several hours (1-8 hours)  & 7 = more than one year\\
      NA = I don't know &
  \end{tabular}
    }\\
\hline
\textit{archiving$^*$} \newline (ARC) \refRQ{2}
& Has the sketch/diagram been archived or will it be archived? 
&  { \renewcommand{\arraystretch}{1}
        \hspace{-8pt}
      \begin{tabular}[t]{llll}
  	0 = no &
	1 = yes, on paper &
	2 = yes, digitally &
	3 = yes, digitally and on paper\\
	& & & NA = I don't know
  \end{tabular}
\newline{\it Furthermore, we asked why or why not the sketch or diagram was archived (open-ended).}
    }\\
\hline
\textit{formality$^*$} \newline (FOR) \refRQ{3}
& 
Please try to specify the formality of your sketch/diagram.
& { \renewcommand{\arraystretch}{1}
        \hspace{-8pt}
     	\begin{tabular}[t]{ll}
	0=very informal { }to{ } 5=very formal & NA = I don't know\\
	(6-point Likert scale item)
	\end{tabular}
	}\\
\hline
\textit{UML$^*$} \newline (UML) \refRQ{3}
& To which degree does the sketch\! /\! diagram contain UML elements?
& { \renewcommand{\arraystretch}{1}
        \hspace{-8pt}
     	\begin{tabular}[t]{ll}
	0=no UML elements { }to{ } 5=only UML elements & NA = I don't know\\
	(6-point Likert scale item)
	\end{tabular}
	}\\
\hline
\textit{purpose} \newline (PUR) \refRQ{4}
& The sketch/diagram helped me to ... (none or multiple answers possible)
& {understand source code / understand an issue / design a new architecture / design new GUI components / design new features / review source code / refactor source code / debug source code / explain source code to someone else / explain an issue to someone else / analyze requirements / support managing the project / other task(s)
}
\\
\hline
\textit{artifacts} \newline (ART) \refRQ{5}
& Please select the software artifact(s) to which the content of the sketch/diagram is related.
  (none or multiple answers are possible)
& (single or multiple) statement(s) or expression(s) /
    (single or multiple) attribute(s), parameter(s), propertie(s), or variable(s) /
    (single or multiple) method(s), function(s), or procedure(s) /
    (single or multiple) classe(s), object(s), or prototype(s) /
    (single or multiple) package(s), namespace(s), module(s), unit(s), or folder(s) /
    (single or multiple) project(s) /
    other artifact(s)\\
\hline
\textit{help-self$^*$} \newline (HES) \refRQ{5}
& Do you think that the sketch/diagram could help you in the future to understand the related source code artifact(s)?
& { \renewcommand{\arraystretch}{1}
        \hspace{-8pt}
     	\begin{tabular}[t]{ll}
	0=it will definitely not help { }to{ } 5=it will definitely help & NA = I don't know\\
	(6-point Likert scale item)
	\end{tabular}
	}\\
\hline
\textit{help-others$^*$} \newline (HEO) \refRQ{5}
& Do you think that the sketch\! / \!diagram could help other software developers in the future to understand the related source code artifact(s)?
& { \renewcommand{\arraystretch}{1}
        \hspace{-8pt}
     	\begin{tabular}[t]{ll}
	0=it will definitely not help { }to{ } 5=it will definitely help & NA = I don't know\\
	(6-point Likert scale item)
	\end{tabular}
	}\\
\hline
%
%
\textit{area} \newline (ARE) \refRQ{4}
& What is the main application area of the project (for which the sketch/diagram was created)?
& software tools / web development / computer games / public media / telecommunications / financial services / health / retail / manufacturing / automotive systems / aerospace / real-time systems / civil service / other\\
\hline
\textit{team-size$^*$} \newline (TES) \refRQ{4}
& How many persons work on this project?
& See variable \textit{contributors}. \\
\hline
\textit{model-driven$^*$} \newline (MDR) \refRQ{4}
& Does the project team employ model-driven software engineering?
& { \renewcommand{\arraystretch}{1}
        \hspace{-8pt}
     	\begin{tabular}[t]{ll}
	0=never { }to{ } 5=always & NA = I don't know\\
	(6-point Likert scale item)
	\end{tabular}
	}\\
\hline
\textit{agile$^*$} \newline (AGI) \refRQ{4}
& To which degree does the team employ agile software development methods?
& { \renewcommand{\arraystretch}{1}
        \hspace{-8pt}
     	\begin{tabular}[t]{ll}
	0=only using agile methods { }to{ } 5=only using other methods & NA = I don't know\\
	(6-point Likert scale item)
	\end{tabular}
	}\\
\hline
\textit{usage$^*$} \newline (USE) \refRQ{1}
& When did you use (look at, modify, extend) the last sketch or diagram that you did not create yourself?
& See variable \textit{creation}. \\
\hline
\textit{gender} \newline (GEN)
& Your gender: 
& male / female \newline (optional) \\ 
\hline
\textit{age$^{**}$} \newline (AGE)
& Your age: 
& 0-99 year(s) \newline (optional) \\ 
\hline
\textit{experience$^{**}$} \newline (EXP)
& Your professional work experience in software development: 
& 0-99 year(s)\newline (optional, please round up to full years.)\\
\hline
\textit{work-time$^{**}$} \newline (TIM)
& How much of your work time is dedicated to software development? 
& 0-100\% \newline (optional) \\
\hline
\textit{occupation} \newline (OCC)
& Your current occupation? 
& none / software developer / software architect / project manager / ... / other \\
\hline
\textit{organization} \newline (ORG)
& What type of organization do you work in?
& government / educational / very small company (<10 employees) / small company (10-50 employees) / medium company (51-1000 employees) / large company (>1000 employees) / self-emplyed\\
\hline
\textit{country} \newline (COU)
& Which country do you work in? 
& country code \newline (Germany=DE, United States of America=US, etc.) \\
\hline
& Additional remarks (open-ended):
& remarks regarding the sketch or diagram used to answer the above questions , the questionnaire as a whole, or the general usage of sketches and diagrams in software development\\
\hline

\end{tabularx}
\label{tab:variables}
\end{table*}

\section{Results}
\label{sec:results}

\begin{table*}[tbp]
\renewcommand{\arraystretch}{1.3}
\tiny
\centering
\caption{Quasi-experiments: Wilcoxon rank-sum test, Spearman's rho, and Cliff's delta. One asterisk indicates that the two-tailed $p$-value is smaller than $0.05$, two asterisks indicate a $p$-value smaller than $0.01$. $CI_d$: confidence interval of Cliff's delta at $95\%$ confidence level.}
\label{tab:results}
\begin{tabularx}{\textwidth}{| ll | Xrrrr | Xrrrr | rrrr |}
\hline
\textbf{G.Var} & \textbf{Var} & \textbf{Group 1} &&&&& \textbf{Group 2} &&&&& \multicolumn{1}{c}{$W$} & \multicolumn{1}{c}{$\rho$} & \multicolumn{1}{c}{$d$} & \multicolumn{1}{c|}{$CI_d$}\\
& & Value(s) & \multicolumn{1}{c}{$n$} & \multicolumn{1}{c}{Mdn} &  \multicolumn{1}{c}{Mod} &  \multicolumn{1}{c|}{IQR} &Value(s) & \multicolumn{1}{c}{$n$} & \multicolumn{1}{c}{Mdn} &  \multicolumn{1}{c}{Mod} &  \multicolumn{1}{c|}{IQR} &&&& \\ 
\hline
\multirow{3}{1cm}{REV} & LSP &  \multirow{3}{1.8cm}{no} & 137 & 3 & 2 & 3 &  \multirow{3}{1.8cm}{yes (once)\newline yes (multiple times)} & 233 & 5 & 6 & 2 &  $7106^{**}$ &  $0.37^{**}$ & $0.43$ & $(0.32, 0.53)$\\
& FOR & & 136 & 1 & 0 & 2 & & 243 & 2 & 1 & 2 &  $43858^{**}$ &  $0.26^{**}$ & $0.30$ & $(0.19, 0.41)$\\  
& ARC & & 130 & 0 & 0 & 1 & & 234 & 1 & 1 & 1 &  $63024^{**}$ &  $0.24^{**}$ & $0.26$ & $(0.15, 0.36)$\\
\hline
\multirow{2}{1cm}{LSP} & EFF & \multirow{2}{1.8cm}{0, 1, 2} & 84 & 0 & 0 & 1 & \multirow{2}{1.8cm}{5, 6, 7} & 189 & 1 & 1 & 1 &  $25095^{**}$ &  $0.51^{**}$ & $0.60$ & $(0.49, 0.69)$\\
& ARC & & 79 & 0 & 0 & 1 & & 179 & 1 & 1 & 0 &  $30358^{*}$ &  $0.49^{**}$ & $0.56$ & $(0.43, 0.66)$\\ 
%
%
%
\hline
\multirow{5}{1cm}{ARC} & LSP &  \multirow{5}{1.8cm}{no} & 139 & 3 & 2 & 2 & \multirow{5}{1.8cm}{yes (paper),\newline yes (digital),\newline yes (both)} & 218 & 6 & 6 & 2 &  $6344^{**}$ &  $0.49^{**}$ & $0.58$ & $(0.48, 0.66)$\\
& EFF & & 141 & 1 & 0 & 1 & &  228 & 1 & 1 & 1 &  $44991^{**}$ &  $0.46^{**}$ & $0.51$ & $(0.41, 0.60)$\\
& HES & & 130 & $2.5$ & 1/3 & 3 & & 220 & 4 & 5 & 2 &  $11940^{**}$ &  $0.41^{**}$ & $0.48$ & $(0.37, 0.58)$\\
& HEO & & 133 & 2 & 3 & 3 & & 222 & 4 & 5 & 2 &  $14671^{**}$ &  $0.41^{**}$ & $0.47$ & $(0.36, 0.57)$\\  
& FOR & & 140 & 1 & 0 & 2 & & 227 & 2 & 1 & 3 &  $39819^{**}$ &  $0.36^{**}$ & $0.42$ & $(0.31, 0.52)$\\ 
\hline
\multirow{2}{1cm}{MED} & CON & \multirow{2}{1.8cm}{trad., int. whiteboard} & \multirow{2}{0.28cm}{77} & \multirow{2}{0.15cm}{2} & \multirow{2}{0.15cm}{2} & \multirow{2}{0.14cm}{1} &  paper & 156 & 1 & 1 & 1 &  $8814^{**}$ & -$0.44^{**}$ & -$0.50$ & (-0.61, -0.36) \\ 
\cline{8-16}
& CON &&&&&& comp., tablet & 157 & 1 & 1 & 1 &  $7929^{**}$ &  -$0.29^{**}$ & -$0.33$ & (-0.46, -0.19)\\  
\hline
\multirow{6}{1cm}{MED} & LSP & \multirow{6}{1.8cm}{paper, trad. whiteboard} & 219 & 4 & 4 & 3 & \multirow{6}{1.8cm}{computer, tablet, int. whiteboard} & 155 & 6 & 6 & 2 &  $5814^{**}$ &  $0.54^{**}$ &  $0.62$ &  $(0.52, 0.70)$\\
& ARC & & 211 & 0 & 0 & 1 & & 157 & 1 & 1 & 0 &  $12403^{**}$ &  $0.53^{**}$ & $0.56$ & $(0.47, 0.64)$\\
& EFF & & 226 & 1 & 1 & 1 & & 161 & 2 & 2 & 1 &  $40763^{**}$ & $0.47^{**}$ & $0.52$ & $(0.42, 0.60)$\\
& FOR & & 228 & 1 & 0 & 2 & & 158 & 3 & 4 & 3 &  $36741^{**}$ &  $0.44^{**}$ & $0.50$ & $(0.40, 0.60)$\\
& UML & & 224 & 0 & 0 & 2 & & 157 & 3 & 0 & 4 &  $47448^{**}$ &  $0.33^{**}$ & $0.37$ & $(0.26, 0.48)$\\ 
& HEO & & 212 & 3 & 3 & 3 & & 158 & 4 & 4 & 2 &  $13219^{**}$ &  $0.30^{**}$ & $0.35$ & $(0.23, 0.45)$\\ 
\hline
\multirow{2}{1cm}{HES} & LSP & \multirow{2}{1.8cm}{0, 1} & 67 & 3 & 2 & 3 & \multirow{2}{1.8cm}{4, 5} & 196 & 6 & 6 & 2 &  $2752^{**}$ &  $0.42^{**}$ & $0.55$ & $(0.42, 0.66)$\\ 
& ARC & & 64 & 0 & 0 & 1 & & 194 & 1 & 1 & 0 &  $18915^{**}$ &  $0.42^{**}$ & $0.50$ & $(0.36, 0.62)$\\
%
%
\hline
\multirow{2}{1cm}{HEO} & LSP & \multirow{2}{1.8cm}{0, 1} & 75 & 3 & 2 & 2 & \multirow{2}{1.8cm}{4, 5} & 179 & 6 & 6 & 2 &  $2321^{**}$ &  $0.50^{**}$ & $0.62$ & $(0.49, 0.72)$ \\ 
& ARC & & 73 & 0 & 0 & 1 & & 180 & 1 & 1 & 0 & $16290^{**}$ &  $0.43^{**}$ & $0.58$ & $(0.48, 0.66)$\\
\hline
\multirow{2}{1cm}{COU} & CRE & \multirow{2}{1.8cm}{DE} & 211 & 3 & 3 & 1 & \multirow{2}{1.8cm}{other countries} & 182 & 3 & 3 & 2 &  $4983^{**}$ &  -$0.07$\hspace{1em} & $0.08$ & $($-$0.04, 0.19)$ \\ 
& USE & & 198 & 3 & 3 & 2 & & 170 & 3 & 3 & 2 & $14535^{**}$ &  -$0.09$\hspace{1em} & $0.10$ & $($-$0.01, 0.22)$\\
\hline
\end{tabularx}
\end{table*}

In this section, we present the methods we used to analyze the survey data, describe the participants, and report on how the data answers our research questions.

\subsection{Methods}
\label{sec:statistical-methods}

We analyzed the responses to the closed-ended questions by means of descriptive statistics and quasi-experiments~\cite{William02}, and the responses to the open-ended questions using open coding~\cite{Corbin08}.
The results of the quasi-experiments are shown in Table~\ref{tab:results}:
The first column indicates the grouping (or quasi-independent) variable, i.e., the variable which was used to divide the responses in two or more groups.
These groups were then analyzed using the variable in the second column (the dependent variable).
Since we used pairwise deletion of missing values and in some cases ignored the middle values of the 6-point Likert scales (2,3), we state the number of responses for each group and variable (\textit{n}).
Furthermore, we provide the median (\textit{Mdn}), the mode (\textit{Mod}), and the interquartile range (\textit{IQR}).

We applied the nonparametric \textit{Wilcoxon rank-sum test} (\textit{W})~\cite{Wilcoxon45} to test if the distributions in the two groups differ significantly. We did not use parametric tests because our variables did not have interval scaling and not all variables were normally distributed.
Likert items, for instance, provide only ordinal data, because the intervals between scale values are not equal~\cite{Jamieson04}.
As shown in the table, all presented group pairs have significantly different distributions (all $p$-values $<0.05$).
We calculated \textit{Spearman's rank correlation coefficient} ($\rho$)~\cite{Spearman04} to test the statistical dependence between two variables.
This coefficient works on ordinal data and does not require normal distribution.
The values range between $+1$ and $-1$.
A positive value indicates a positive correlation, i.e., if one variable increases, so does the other; a negative value of $\rho$ indicates a negative correlation, i.e., the variables vary in opposite directions.
Our interpretation of the values of $\rho$ is based on the following scheme: weak ($0.1\leq|\rho|<0.3$), moderate ($0.3\leq|\rho|<0.5$), and strong ($0.5\leq|\rho|\leq 1$), which is derived from Cohen's definitions \cite{Cohen88}.
Apart from a few exceptions, we only considered results having at least a moderate correlation.
To measure the effect size, we used \textit{Cliff's delta} ($d$)~\cite{Cliff93}.
Its values range between $+1$, when all values of the second group were higher than the values of the first group, and $-1$, when the reverse was true.
Moreover, we provide the confidence interval of $d$ at a 95\% confidence level.

The qualitative data was generated by the three open-ended questions in our questionnaire.
In total, 343 respondents (87\%) answered to the questions why or why not they archived their sketch.
Furthermore, we received 69 general remarks (18\%) with diverse opinions on the respondents' usage of sketches and diagrams.
We analyzed the answers using open coding~\cite{Corbin08} and assigned them to categories.
In the following, we will refer to statements made by participants in the open-ended questions using their ID (e.g., P12 meaning the participant with ID 12).

\subsection{Survey Participants}

Overall, 394 persons (361 male, 11 female, 22 unknown) with a median age of 34 filled in our questionnaire.
Of the participants that indicated their age, 74\% were between 20 and 40 years old and 24\% were older than 40, but younger than 60.
The respondents worked in 32 different countries, most of them in Germany (54\%) or North America (15\%). 

52\% of our respondents worked as software developers, 22\% as software architects.
The rest included project managers (5\%), consultants (5\%), industrial as well as academic researchers (6\%), and students (5\%).
86\% of them spent most of their work time developing software; the median value was 80\%.
47\% had more than 10 years of professional work experience, while 21\% had less than 5 years.
The median professional work experience was 10 years.
The respondents worked with companies of very different sizes (27\% with up to 50 employees and 29\% with more than 1000 employees) and the application areas of their projects included software tools, web development, financial services, automotive, manufacturing, and health, to name a few.

Since over half of our participants came from Germany, we were interested if their answers were consistent with the answers of non-German participants.  
To this end, we employed quasi-experiments to compare these groups and found, beside demographic data, no major differences (see Table \ref{tab:results} for the results of this test for variables \textit{creation} and \textit{usage}).

\subsection{Findings on Research Questions}

With the results of the quasi-experiments, the qualitative data, and descriptive statistics, we can now answer our research questions.
When using data from the table, we provide the values of the first two columns to identify the row we are referring to, e.g., \refTab{REV}{FOR} refers to the second row.

\subsubsection{Creation, Usage, Effort, and Revision (RQ1)}

To assess the frequency of sketch creation, we asked the respondents when they created their last sketch.
24\% created their last sketch on the same day, another 39\% within a week, another 22\% within a month, and another 14\% created their last sketch more than one month ago.
Hence, 64\% of the sketches were created at most several days ago.
We also asked the respondents about the last time they used (looked at, modified, extended) a sketch that was not created by themselves.
27\% used it on the same day, 34\% within a week, 17\% within a month, and another 15\% used it more than one month ago.
Thus, 61\% of the respondents used a sketch made by someone else at most several days ago.
Overall, most respondents ($77\%$) created and/or used sketches within the last week.

To assess the effort of creating a sketch, we asked the respondents to estimate the effective work time that went into the creation and revision of the sketch. If several persons were involved, we asked them to add up their individual work times. More than two thirds ($68\%$) of the sketches were created in less than one hour, 25\% were created in several hours and in only four cases, the creation of the sketch took more than five days.

After creation, about 15\% of the sketches were revised once, and 47\% multiple times.
73\% of the sketches that were not revised were created on analog media, compared to 49\% of the revised ones.
The median lifespan of revised sketches was several weeks, whereas the median lifespan of non--revised sketches was several hours \refTab{REV}{LSP}.
Revised sketches were also more likely to be archived \refTab{REV}{ARC} and less informal \refTab{REV}{FOR}.

It was common that people created a new version of an outdated sketch, extended an existing one, or just captured their analog whiteboard drawings.
P75 named an example for the latter: He wrote that he and his team ``always take a photograph of the sketch (we all have smartphones!) and email the photo to the team members and place it in a shared wiki as well''.
Another transition was described by P193, who wrote that he ``started with a whiteboard drawing, then a more detailed pencil\&paper sketch, and eventually it was modeled in yEd.''
Similar workflows were reported by P41, P52, and P149.

Sketches were not only redrawn, but were also transcribed to other representations. 
P173 noted that his sketch will be ``replaced by textual documentation'', similar to P222, who speaks of ``formalization [...] in text form'', meaning source code comments.
A sketch may also be replaced simply by a mental model that the creator built with the help of the sketch (see also \cite{LaToza06}): P373, for instance, wrote that his sketch supported ``knowledge transfer from explicit (paper) to tacit (in head) knowledge''.
Recreating sketches digitally for documentation seems to be a common use case.
P247 mentioned that his sketch ``has been transformed digitally by using a sketch tool and added to the development documentation''.
Similar workflows were, for instance, reported by P21, P23, and P290. 
Once the sketches were digital, it was easier to update them. P305 wrote about his digital sketch that ``over the next several months we will be working from it and changing it as we learn''.
P89 stated that he always starts ``a new project with a diagram, which is modified as work progresses''.

\begin{normalbox}
Creating and using sketches are frequent tasks among software practitioners.
Most sketches were created in less than one hour and more than half of them have been revised.
Transitions between different media were common.
\end{normalbox}

\subsubsection{Lifespan and Archiving (RQ2)}
\label{sec:lsp-arc}

We asked the participants to estimate the lifespan of their sketch, i.e., how long they did or will use it. 21\% used it for less than one hour, 9\% at most for one day, 32\% for less than one month, and another 33\% for one month or more.
The median lifespan was several weeks.
Less effort went into the creation of sketches with a short estimated lifespan \refTab{LSP}{EFF}.
These sketches were also less likely to be archived than those with a longer estimated lifespan \refTab{LSP}{ARC}.

Overall, more than 58\% of the sketches were archived (6\% only on paper, 42\% digitally, and 10\% both, digitally and on paper).
Almost all digital sketches were archived (94\%), but also 38\% of the analog ones.
More effort went into the creation of archived sketches \refTab{ARC}{EFF} and the lifespan of those sketches was estimated to be several months---compared to several hours for sketches that were not archived \refTab{ARC}{LSP}.
Archived sketches were more formal \refTab{ARC}{FOR} and would more likely help the respondent \refTab{ARC}{HES} or others \refTab{ARC}{HEO} to understand the related source code artifact(s) in the future.

We asked the respondents to comment on why or why not they archived their sketch.
We categorized the answers to both questions independently and identified nine categories in total:
Four categories indicating why a sketch was archived, four categories indicating why not, and one category for all answers with hints on the general archiving practice.
Please note that the categories are not disjoint and have different granularity. One answer may belong to several categories.

\myCaption{Reasons for archiving a sketch}

To the first category, we assigned answers indicating that the sketch or diagram was kept as \textit{documentation}.
The majority of answers in this category pointed out that the sketch documents the implementation, e.g., the architecture, structure, states, or data flows.
Many respondents explicitly mentioned the documentation of source code artifacts like APIs, components, or test cases.
Some of them reported on the documentation of requirements and specifications, decisions, ideas, solutions, or discussions.
P365 archived his sketch because it ``document[s] the discussion'' and it will be used to ``further investigate into [the] sketched idea''.
However, sketches may also document mistakes.
P327 wrote that he archived his sketch to ``trace the cause of decision'' and that the sketch ``could be useful to explain our mistake later''.
P369 stated that he posted his diagram on the wall of his office, because ``charts and diagrams document where you have been, what [you] were [...] thinking, and where [...] you intend to go''.

To the second category, we added answers pointing out that the sketch was or will be archived for \textit{future use}.
A common reason for archiving a sketch was to be able to reference it in the future.
Other future uses included reusing the sketch as a template, as well as updating, refining, expanding, or digitally recreating it.
Furthermore, sketches were used for planning, bug fixing, as a reminder, or for communication with customers or team members (e.g., as input for a discussion).
Sketches and diagrams were also used during implementation or for code and interface generation.
Several respondents stated that they archived their sketch to be able to explain parts of the software system or ideas to other stakeholders (e.g., for onboarding new staff).

The visual artifacts in the third category support \textit{understanding} and were thus archived.
The answers showed that understanding the implementation was a central aspect.
Some respondents archived their sketch because it helped them to understand the whole project, others named the understanding of processes, workflows, problems, ideas, or decisions.
P162 wrote that his sketch ``greatly aids in understanding the basic architectural concepts''.
P340 even stated that ``it will be difficult to understand the code'' without the diagram.
An interesting remark by P233 highlights the connection of sketches and ideation.
He wrote that he keeps sketches ``as a personal archive of knowledge and ideas''.

The fourth category, named \textit{visualization}, is closely related to the previous one.
However, many participants explicitly mentioned that they archived their sketch because it visualizes a process, problems, requirements, software, or other concepts.
Some mentioned that they prefer visual representations of software over text. P21, for instance, wrote that his sketch ``is stored [...] in case I or someone else analyzes the sketched part of code. This way, it can be quickly understood due to the visual representation without hours of digging through complex source code''.
P145 noted that his sketch ''explains a data flow better than in spoken words''.
P162 even states that his sketch ``shows concepts that are not directly visible from code''. 
Sketches reduce the cognitive load, as P85 reported: ``I generally use them to visualize a process that I can't keep in my head all at once [...].''
The team of P65 uses flip chart sketches for code reviews, because ``it helps to get a grasp on the structure and make the code concrete and available to the involved persons''.

\myCaption{Reasons for not archiving a sketch}

The main reason why respondents did not archive their sketch was that they thought it \textit{served its purpose} and, thus, was not worth keeping. 
The named purposes included understanding, explaining, visualizing, designing, communicating, prototyping, problem solving, and structuring thoughts, ideas, or the implementation.
P2, for instance, wrote that he did not archive his sketch because he ``just created it to visualize a very special situation in the work flow of the program''.
P6 mentioned that he used his sketch to ``think through a problem'' and it had ``no value as a long-term reference''.
Interesting is the fact that sketches were also created where documentation was available: P4 reported that his sketch served his ``own understanding of a well-documented system''.

The second category of answers indicated that the sketch was \textit{substituted} by another representation, e.g., being replaced by a new or extended version, redrawn on another medium, or implemented in source code.
P204 stated that his sketch is ``useless after implementing the ideas of the sketch into source code''.
This was a common reason for not keeping a sketch, as it was reported by several participants (e.g., P7, P47, P217, P256, and P334). 
P332 wrote that ``the code will be the final representation of the idea, the sketch is just scaffolding''.
To the third category, we added answers indicating that the sketch was \textit{outdated}.
The main reason for this was the evolution of the related software.
P27, for instance, noted that ``the software will be developed further and diverge from the sketch over time''.
Similar situations were reported by P39, P52, P102, and P339.
In the fourth category, we summarized all answers that named some kind of \textit{technical issue} as a reason for not keeping the sketch.
P78, for instance, wanted to keep his sketch, but wrote that he had ``no way to archive whiteboard drawings''.
P123 reported that his sketch ``ended up in code'' and ``there is no good option to keep it together [with source code]''.
Similarly, P2 wrote that ``in case there was an easy way to combine both, code [...] and sketch I might have thought about archiving it''.
P259 noted that ``there is no special place where to archive'' the sketch and he also addressed the issue that ``nobody would update it, if the software artifacts change''.
Another problem is that contextual information may be necessary to understand a sketch, as reported by P314.

\myCaption{Archiving practice}

To this category, we added all answers that contained hints to the respondent's general \textit{archiving practice} or the systems used for storage.
Sketches were stored, for instance, in wikis, version control systems, issue tracking systems, or emails.
Some participants reported that they try to archive as much as possible, like P10, who stated that ``every artifact in the process of creating a software should be archived''. 
Four participants named compliance or regulatory demands as a reason for keeping their sketch.

\begin{normalbox}
One third of the sketches had an estimated lifespan of one day or less, one third of up to one month, and another third of more than one month.
The majority of sketches were archived, most of them digitally. 
Many sketches were kept because they document or visualize parts of the implementation or assist its understanding.
\end{normalbox}

\subsubsection{Formality and UML (RQ3)}

\begin{figure}[h!]
\centering
\includegraphics[width=\columnwidth]{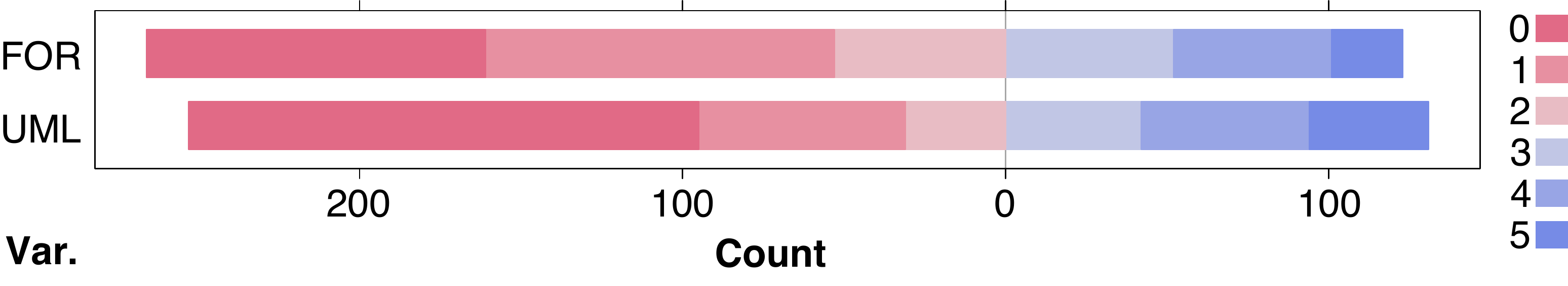}
\vspace{-2\baselineskip}
\caption{Distribution of answers for FOR and UML}
\label{fig:dist-for-uml}
\end{figure}

In the questionnaire, we defined formality as the degree to which a sketch follows a pre-defined visual language (e.g., the UML). 
68\% of the sketches were rated as informal (Likert 0-2).
While 27\% rated their sketch as very informal (Likert 0) only 6\% found their sketch to be very formal (Likert 5).
We also explicitly asked the respondents about the use of UML elements in their sketches.
While 40\% of the sketches contained no UML elements at all, 9\% consisted solely of UML elements. 
Overall, 24\% found that their sketch contained few UML elements (Likert 1-2); another 24\% found that their sketch contained more UML elements (Likert 3-4).
However, 30\% of the sketches that contained more UML elements were still rated as informal (Likert 0-2).
See Figure \ref{fig:dist-for-uml} for a diverging stacked bar chart~\cite{Robbins11} of the answers for the variables \textit{formality} and \textit{UML}.

In total, 18 of the respondents' general remarks were about their use of UML or their general opinion on such formal notations.
The opinions ranged from completely rejecting formal methods (P83) to very positive ones (P194).
One argument against UML or other formal notations was that ``most of the time, you'd have to read the code anyways'' (P8).
P102 states that ``UML is often not known, and almost never used''.
According to him, ``people prefer to code or to get code (even buggy) rather than to draw little drawings''.
On the other hand, P194 stated that he thinks that diagrams ``help a lot in designing good object-oriented systems''.
P210 stated that he prefers having less text documentation and more diagrams, because ``people tend not to read written specifications accurately but spend more time understanding a diagram''.
Most of the remarks indicated an informal sketching practice, meaning that if UML was used, it was not used strictly as defined in the standard (e.g., reported by P21, P94, P190, and P304).
This is in accord with our qualitative results described above. 
Informal UML usage also influenced the choice of medium, as P210 reported that he creates most of his sketches on paper, because they ``combine UML, icons and mind-mapping elements, as well as graphical sketches to visualize functionality''.

\begin{normalbox}
The majority of sketches and diagrams were informal.
Whereas 40\% of them contained no UML elements at all, 48\% contained at least some, and only 9\% consisted solely of UML elements.
Respondents' remarks indicate that if UML is used, it is often not used strictly as defined in the standard. 
\end{normalbox}

\subsubsection{Media, Context, and Purpose (RQ4)}

Almost 60\% of the sketches were drawn on analog media like paper (40\%) or traditional whiteboards (18\%).
The remaining sketches were almost exclusively drawn on computers (39\%).
Only five sketches were drawn on an interactive whiteboard and only three on tablets or smartphones.
Sketches created on paper or digitally were most likely created alone, whereas sketching on traditional whiteboards was more likely to be done collaboratively \refTab{MED}{CON}.
The medium and lifespan of a sketch were also related: Sketches created on analog media (paper or traditional whiteboards) had an estimated lifespan of several work days, whereas sketches created digitally (computer, tablet, or smartphone) had an estimated lifespan of several months \refTab{MED}{LSP}.
Furthermore, digital sketches were more likely to be archived than analog ones \refTab{MED}{ARC}, they were more formal {\scriptsize (MED, }{\scriptsize FOR)}, and were more likely to contain UML elements {\scriptsize (MED, }{\scriptsize UML)}.
Digital sketches were also rated as being more likely to help others in the future to understand the related source code artifact(s) \refTab{MED}{HEO}.
Besides, more effort was put into digital sketches \refTab{MED}{EFF}.

While 51\% of the sketches were created by a single person, 28\% by two persons, and 15\% by three persons, only 6\% were created by more than three persons. Actually, only one sketch was created by more than 10 persons. 
Sketches to which one or two people contributed were most likely created on paper (46\%).
38\% of these sketches were created using a computer and 15\% on a traditional whiteboard.
When more than two people contributed to the sketch, the ratio of computer sketches increases only slightly (43\%).
However, the ratio of traditional whiteboards doubles (33\%) and the ratio of paper sketches halves (17\%).

\begin{normalbox}
Most sketches were drawn on analog media like paper or whiteboards.
Half of them were created by a single person and another third by two persons---only few were created by more than three persons.
Paper was predominantly used alone, whiteboards collaboratively.
\end{normalbox}

To capture the creation context, we asked for the application area, team size, and employment of model-driven software engineering or agile methods in the software project for which the sketch was created.
The most common application areas were software tools (27\%), followed by web development (18\%), and financial services (11\%).
In 54\% of the cases, the project teams never or only rarely employed model-driven software engineering (Likert 0-1), whereas 42\% of the teams intensively employed agile methods (Likert 4-5).
As mentioned before, the respondents worked with companies of very different sizes.
The most common team size was  4 to 10 (40\%), 11\% of the respondents worked alone, 8\% with one colleague, and 19\% with two colleagues. 
15\% worked in teams of 11 to 50 people, whereas only 5\% worked in teams with more than 50 employees.
However, we found no significant influence of the team context on the sketching behavior of our participants.

We also asked the respondents about the purpose of their sketch.
They could choose multiple answers from a given list of tasks that the sketch helped to accomplish (see Table \ref{tab:variables}).
The most frequent tasks were designing a new architecture (52\%), designing new features (48\%), explaining an issue to someone else (46\%), analyzing requirements (45\%), and understanding an issue (44\%); the least common tasks were reviewing source code (9\%) and debugging source code (7\%).
Overall, most tasks were either related to designing (75\%), explaining (60\%), or understanding (56\%).
Of the most frequent tasks only analyzing requirements cannot be assigned to these categories.

Three participants mentioned sketches as a medium to outline the high-level system architecture (P8, P26, P304).
P374 thinks that ``sketches and diagrams are critical to understanding software projects and architectures''.
Sketching is also used to communicate with clients (P106, P112) or between ``business and development'' (P224).

\begin{normalbox}
The most common purposes for creating sketches were related to designing, explaining, or understanding.
Furthermore, analyzing requirements played an important role.
\end{normalbox}

\subsubsection{Relation to Source Code and Value (RQ5)}

We asked the respondents to select the software artifact(s) to which the content of their sketch was related. 
They could choose multiple answers from a given list of artifacts, which was sorted in order of increasing level of abstraction (e.g., statement, method, class, package).
For each level, we not only named terms used in object-oriented programming, but also similar concepts used in other paradigms (see Table~\ref{tab:variables}).
Furthermore, respondents could indicate whether the sketch was related to a single instance or to multiple instances of the chosen artifact.
For the sake of brevity, we will only name one representative of each level and do not distinguish single and multiple instances in the following.
If a sketch is related to a lower level of abstraction this normally implies that it is also related to the levels above.
Thus, we base our interpretation on the most specific artifact the participant selected, i.e., the artifact with the lowest level of abstraction.
9\% of the sketches were most specifically related to statements, 8\% to attributes, 20\% to methods, 23\% to classes, 17\% to packages, and 19\% to projects.
We can conclude that sketches rarely pertain to certain attributes or statements, but rather to methods, classes, packages, or projects.

We asked the respondents to assess if their sketch could help them or others to understand the related source code artifact(s) in the future.
52\% of the sketches were rated as helpful (Likert 4-5) for the respondent, 47\% were rated as helpful for others.
Helpful sketches had a longer estimated lifespan \refTab{HES}{LSP} \refTab{HEO}{LSP} and were more likely to be archived than not helpful sketches \refTab{HES}{ARC} \refTab{HEO}{ARC}.

\begin{normalbox}
Sketches were rarely related to certain attributes or statements, but rather to methods, classes, packages, or projects.
About half of the sketches were rated as helpful to understand the related source code artifact(s) in the future.
 \end{normalbox}
 
\subsection{Correlations}
\label{sec:correlations}

To estimate the strength of the correlations between the captured variables we used \textit{Spearman's rank correlation coefficient}  $\rho$ (see Section \ref{sec:statistical-methods}).
To test the significance of each correlation coefficient, we computed the two-tailed $p$-values and checked whether these $p$-values were both less than $\alpha=0.05$ and $\alpha=0.05/15$ (Bonferroni correction~\cite{Dunn61}, as we computed a total of 15 correlations).  
Since we didn't start with apriori hypotheses about the correlations, we only consider moderate and strong correlations that are significant after Bonferroni correction.

The results are shown in Table~\ref{tab:correlations}:
The strongest correlation was found between \textit{formality} and \textit{UML}, other strong correlations exist between \textit{archiving} and \textit{lifespan}, and between \textit{effort} and \textit{lifespan}.
Furthermore, all pairwise correlations of the four variables \textit{formality}, \textit{archiving}, \textit{effort}, \textit{lifespan} and the three variables \textit{effort}, \textit{lifespan}, \textit{revision} are at least moderate. 
Note that one has to be cautious when interpreting these correlations, because correlation does not imply causality.
For example, the above mentioned correlations of the variable \textit{effort} can be put in other words:
The more effort is put into a sketch, the more likely it is that it will be used for a longer time, that it will be archived, and that it will be more formal.
But, for example, we may not conclude that effort is the cause for archiving a sketch.

As mentioned above, the four variables \textit{formality}, \textit{archiving}, \textit{effort}, and \textit{lifespan} each correlate at least moderately.
We analyzed the open-ended answers and searched the data for sketches illustrating these correlations:
P156 created his rather formal sketch (formality: 3) using a traditional whiteboard and archived it digitally.
He spent several hours creating it and estimated its lifespan to be more than one year, noting that it is a ``general architecture sketch'' that ``will help others understand communication and probably won't change in the forthcoming months''.
Another example is the formal sketch (formality: 5) of P122 that was created in more than 5 work days.
It was a ``state diagram'' created digitally that is ``needed as long as the program exists'' (lifespan: more than one year). 
The above examples are sketches that have high values for the considered variables.
However, there are also examples for sketches at the other end of the spectrum.
P73 created his informal sketch (formality: 0) in less than ten minutes on paper.
The sketch had an estimated lifespan of only several minutes and was not archived, since it was a ``temporary sketch for debugging''.
The sketch of P80 had similar characteristics, being a ``transient diagram, used to explain an approach to a problem''. 

\begin{normalbox}
Formality, archiving, effort, and lifespan each correlate at least moderately.
This is also the case for effort, lifespan, and revision.
\end{normalbox}

\begin{table}[tb]
\renewcommand{\arraystretch}{1.3}
\scriptsize
\centering
\caption{Correlation table with Spearman's correlation coefficients $\rho$ (one asterisk: significant at the 0.05 level, two asterisks: remains significant after Bonferroni correction).}
\label{tab:correlations}
\begin{tabular}{c | cccccc}
\hline
$\rho$ &UML & FOR & ARC & EFF & LSP & REV \\
\hline
UML & - & & & & & \\
FOR & $\boldsymbol{0.62^{**}}$ & - & & & &  \\
ARC & $0.26^{**}$ & $\boldsymbol{0.37^{**}}$ & - & & &  \\
EFF & $0.27^{**}$ & $\boldsymbol{0.39^{**}}$ & $\boldsymbol{0.47^{**}}$ & - & &  \\
LSP & $0.26^{**}$ & $\boldsymbol{0.38^{**}}$ & $\boldsymbol{0.52^{**}}$ & $\boldsymbol{0.50^{**}}$ & - &  \\
REV & $0.17^{*}$ & $0.26^{**}$ & $0.24^{*}$ & $\boldsymbol{0.45^{**}}$ & $\boldsymbol{0.36^{**}}$ & - \\
\hline
n & 384 & 389 & 371 & 390 & 377 & 383 \\
\hline
\end{tabular}
\vspace{-\baselineskip}
\end{table}

\section{Discussion}

Creating sketches and using sketches created by others are common tasks among software practitioners.
Most sketches were revised multiple times and had an estimated lifespan of more than one week.
This is backed up by qualitative data, as respondents reported that sketches were often revised or redrawn, especially when used for documentation.
Qualitative data also indicates that it is common that the lifecycle of a sketch starts on analog media like paper or whiteboards and eventually ends as an archived digital version.

More than two thirds of the sketches were created in less than one hour, mostly using analog media like paper or whiteboards. 
This may be a reason for many sketches being rated as informal.
Interactive whiteboards and tablets were almost never used.
The use of UML elements was higher than we expected after the field study, just like the lifespan.
Another unexpected result was that most sketches were archived.
Besides being archived, many sketches were rated as helpful to understand the related source code artifacts in the future. In the open-ended answers, many respondents stated that their sketch helped them to understand issues or implementation details.
Thus, despite the difficulty of keeping them up to date, sketches are a valuable resource for developers.

With our survey, we could validate insights from the interviews we made during our exploratory research:
One interviewed developer noted that in his team, whiteboards are normally used as soon as more than two persons are involved. Otherwise, the preferred medium is paper.
We found out that paper was the prevalent medium for one or two contributors and that whiteboard usage doubled when more than two persons contributed.
Furthermore, we observed that while half of the sketches were created by a single person, the other half was almost entirely created by two or three persons; only few sketches were created by more than three persons.
Our preliminary assumption that sketches and diagrams primarily relate to classes and methods---or other source code artifacts with the same level of abstraction---was partly confirmed, but projects and packages played also an important role.

\subsection{Threats to Validity}

During the qualitative analysis, we tried to mitigate the ``lone researcher bias'' \cite{Burnard08} by applying multiple coding \cite{Barbour01}:
Two researchers performed the coding independently, before they discussed the results and agreed on common categories.
Furthermore, most of the quantitative results were computed independently by both researchers. 
Nevertheless, there exist possible threats to external and construct validity, which we address in the following.

\textit{External validity:} It is obvious that drawing a random sample from the population of software practitioners was impossible for us.
We had to rely on available subjects, which is known as \textit{convenience sampling}~\cite{Gravetter12, Babbie10} or \textit{opportunity sampling}~\cite{Searle00}.
Since we asked people to spread the link to our online questionnaire, we also applied a kind of \textit{snowball sampling}~\cite{Babbie10}.
We had only little control over the representativeness of our sample, because the participants were selected by the channels we used to contact them.
However, we tried to mitigate this threat to external validity by describing our sample in detail to ensure comparability to results of other studies and available demographic data about the target population.
Furthermore, we named the channels we used to contact participants. 

\textit{Construct validity:} In our field study, we explicitly asked about the usual frequency of sketching.
In hindsight, we found that this question ignored that the sketching frequency most likely varies with the different software development phases and it may thus be difficult for the participants to give such a general estimate.
To reduce biases like social desirability and frequency illusion in our survey, we decided not to ask participants about their typical sketching behavior, e.g., how often they sketch or whether they use UML notation, but instead asked them about a concrete artifact, namely their last sketch.
However, a threat to construct validity may be the way we tried to capture the context in which the sketch was created.
Beside asking for the team size and application area of the project, we asked whether the project team ``employs model-driven software engineering''.
This question may be too vague, as participants may have a differing notion of what model-driven software engineering is.
In hindsight, we may have better asked for more concrete tools or practices, e.g. which programming languages are used.
Furthermore, the perception of what exactly ``UML elements'' are may differ between respondents.

For most questions, we asked the participants to think of the last sketch or diagram they created.
Depending on how long the period of time between creation and filling out the questionnaire was, a recall bias may affect the answers. 
As 85\% of our participants created their last sketch not longer than one month ago, this bias is unlikely to influence our results.
Moreover, we cannot rule out the possibility that we may have missed an important confounding variable.
However, since our sketch dimensions are based on past studies and our preliminary research, we think that this is unlikely. 

\subsection{Related Work}
\label{sec:related-work}

Over the past years, studies have shown the importance of sketches and diagrams in software development.
However, sketching is also an important task in other domains.

\textit{Sketches in general:} Artists sketch to clarify existing ideas and to develop new ones~\cite{Fish90}.
In mechanical design, sketches not only document final designs, but also provide designers with a memory extension to help ideas taking shape and to communicate concepts to colleagues~\cite{Ullman90}.
Beside sketches being an external representation of memory and a means for communication~\cite{Tversky02, Tversky01}, they serve as documentation~\cite{Schuetze03}.
Sch\"utze et al. showed that the possibility to sketch has a positive effect on the quality of solutions in early stages of the design process~\cite{Schuetze03}.
Furthermore, the ambiguity in sketches is a source of creativity~\cite{Goldschmidt03, Suwa00, Tversky03} and they support problem-solving and understanding~\cite{Suwa02}.
In our survey, we found that the latter was one of the main reasons why participants archived their sketch. 

\textit{Sketches in software engineering:} Software designers not only use sketches to design the appearance, but also the behavior of software~\cite{Myers08}.
A study of Brown et al.~\cite{Brown08} revealed the importance of sketches for collaboration between user interaction designers and software developers.
Chung et al.~\cite{Chung10} showed that diagramming in distributed environments like open-source projects differs from diagramming in co-located settings.
Dekel and Herbsleb~\cite{Dekel07} studied software design teams at the \textit{OOPSLA DesignFest}, observing that teams intentionally improvise representations to fulfill ad-hoc needs, which arise during the evolution of object-oriented design, and thus diverge from standard-notations like UML.
We can support this with our survey, since most sketches were informal, but often contained at least some UML elements.

Walny et al.~\cite{Walny11-1} analyzed eight workflows of software developers in an academic setting.
They report on a variety of transitions that sketches and diagrams undergo.
Our quantitative as well as qualitative results provided insights into the transitions of sketches.
More than half of the sketches were revised. Furthermore, respondents reported on sketches being shared with others or redrawn digitally.

In another study, Walny et al.~\cite{Walny11-2} analyzed 82 whiteboard drawings in a research institution to achieve a better understanding of what they called spontaneous visualizations.
Our study suggests that one reason for archiving a sketch is that it helps to visualize the implementation, issues, or processes.

LaToza et al.~\cite{LaToza06} conducted a survey with 280 software engineers at Microsoft. They found that paper and whiteboards were perceived as most effective for designing. Furthermore, they state that understanding the rationale behind code is the biggest problem for developers. 
In our study, over half of the sketches helped the respondents to understand source code or general issues.

Cherubini et al.~\cite{Cherubini07} interviewed eight software developers at Microsoft, identifying nine scenarios in which developers created and used drawings.
They explored these scenarios using a survey with 427 participants, also recruited at Microsoft.
We based our list of purposes for sketch creation on their scenarios (see Table~\ref{tab:results}) and found similar results.
However, we asked for further purposes and found analyzing requirements to be an important task.
Cherubini et al.\ state that the use of formal notations like UML was very low.
This is consistent with Petre~\cite{Petre14}, who reports on a series of semi-structured interviews with 50 professional software developers on their UML use. She states that the majority of interviewed practitioners did not use UML and those using UML, tended to use it informally and selectively.
Our study confirms the informal use of UML, but we found that 57\% of the sketches contained at least some UML elements.

\section{Conclusion}

The main contribution of this paper is a thorough description of the manifold dimensions of sketches and diagrams in software development by presenting quantitative as well as qualitative results from a survey with 394 participants.
This survey revealed that sketches and diagrams, even if they are often informal, are a valuable resource, documenting many aspects of the software development workflow.
We showed that sketches are related to different source code artifacts and that roughly half of the sketches were rated as either helpful for the respondents or others to understand these artifacts.
Furthermore, the qualitative data showed that sketches often document or visualize the implementation and support people in understanding it.

As documentation is frequently poorly written and out of date~\cite{Forward02, Lethbridge03}, sketches could fill in this gap and serve as a supplement to conventional documentation like source code comments.
Tool support is needed to assist developers in archiving and retrieving sketches related to certain source code artifacts.
Since more than half of the sketches analyzed in our survey were archived either digitally or digitally and on paper, software professionals are willing to keep their visual artifacts.
However, they also named technical issues, e.g., that there is no good technique to keep sketches together with source code.
A tool should support the evolution of sketches, because more than 60\% of them were revised once or multiple times.
Qualitative data indicates that it is a common use case for sketches to be initially created on analog media like paper or whiteboards and then, potentially after some revisions, they end up as an archived digital sketch. 

Our work is a good starting point for future research:
We plan to evaluate what distinguishes helpful from not helpful sketches and what contextual information is needed to understand them later.
Moreover, it would be interesting to examine if visualizations for certain source code artifacts share common characteristics.
This research may lead to recommendations for software practitioners on how to augment or annotate their sketches so that they can serve as a valuable software documentation.

\subsection*{Acknowledgments}
We want to thank the participants of our survey as well as the software developers we
interviewed and who shared their sketches with us. Thanks to Fabian Beck for his helpful 
suggestions on this paper.  

\clearpage
\balance
%
\bibliographystyle{abbrv}
\bibliography{bibliography}  

\begin{thebibliography}{10}

\bibitem{Babbie10}
E.~Babbie.
\newblock {\em The practice of social research}.
\newblock Cengage Learning, 2010.

\bibitem{ST14}
S.~Baltes and S.~Diehl.
\newblock Sketches and diagrams in practice -- survey data.
\newblock \url{http://www.st.uni-trier.de/survey-sketches/}.

\bibitem{Barbour01}
R.~S. Barbour.
\newblock Checklists for improving rigour in qualitative research: a case of
  the tail wagging the dog?
\newblock {\em British Medical Journal}, 322(7294):1115, 2001.

\bibitem{Brown08}
J.~Brown, G.~Lindgaard, and R.~Biddle.
\newblock Stories, sketches, and lists: Developers and interaction designers
  interacting through artefacts.
\newblock In {\em {AGILE '08: Proceedings of the 11th AGILE Conference}}, pages
  39--50. IEEE, 2008.

\bibitem{Burnard08}
P.~Burnard, P.~Gill, K.~Stewart, E.~Treasure, and B.~Chadwick.
\newblock Analysing and presenting qualitative data.
\newblock {\em {British Dental Journal}}, 204(8):429--432, 2008.

\bibitem{Cherubini07}
M.~Cherubini, G.~Venolia, R.~DeLine, and A.~J. Ko.
\newblock Let's go to the whiteboard: how and why software developers use
  drawings.
\newblock In {\em {CHI '07: Proceedings of the SIGCHI Conference on Human
  Factors in Computing Systems}}, pages 557--566. {ACM}, 2007.

\bibitem{Chung10}
E.~Chung, C.~Jensen, K.~Yatani, V.~Kuechler, and K.~N. Truong.
\newblock Sketching and drawing in the design of open source software.
\newblock In {\em {VL/HCC '10: Proceedings of the IEEE Symposium on Visual
  Languages and Human-Centric Computing}}, pages 195--202. IEEE, 2010.

\bibitem{Cliff93}
N.~Cliff.
\newblock Dominance statistics: Ordinal analyses to answer ordinal questions.
\newblock {\em Psychological Bulletin}, 114(3):494, 1993.

\bibitem{Cohen88}
J.~Cohen.
\newblock {\em Statistical power analysis for the behavioral sciences}.
\newblock Psychology Press, 2nd edition, 1988.

\bibitem{Corbin08}
J.~Corbin and A.~Strauss.
\newblock {\em Basics of qualitative research}.
\newblock Sage Publications, 3rd edition, 2008.

\bibitem{Dekel07}
U.~Dekel and J.~D. Herbsleb.
\newblock Notation and representation in collaborative object-oriented design:
  an observational study.
\newblock In {\em {OOPSLA '07: Proceedings of the 22nd Annual ACM SIGPLAN
  Conference on Object-Oriented Programming, Systems, Languages, and
  Applications}}, pages 261--280. {ACM}, 2007.

\bibitem{Dillman93}
D.~A. Dillman, M.~D. Sinclair, and J.~R. Clark.
\newblock Effects of questionnaire length, respondent-friendly design, and a
  difficult question on response rates for occupant-addressed census mail
  surveys.
\newblock {\em Public Opinion Quarterly}, 57(3):289--304, 1993.

\bibitem{Dunn61}
O.~J. Dunn.
\newblock Multiple comparisons among means.
\newblock {\em Journal of the American Statistical Association},
  56(293):52--64, 1961.

\bibitem{Evans00}
S.~Evans, A.~Kent, and B.~Selic.
\newblock {UML 2000--The Unified Modeling Language}.
\newblock {\em {LNCS number 1939}}, 2000.

\bibitem{Fish90}
J.~Fish and S.~Scrivener.
\newblock Amplifying the mind's eye: sketching and visual cognition.
\newblock {\em Leonardo}, pages 117--126, 1990.

\bibitem{Forward02}
A.~Forward and T.~C. Lethbridge.
\newblock The relevance of software documentation, tools and technologies: a
  survey.
\newblock In {\em {DocEng '02: Proceedings of the 2002 ACM Symposium on
  Document Engineering}}, pages 26--33. ACM, 2002.

\bibitem{Goldschmidt03}
G.~Goldschmidt.
\newblock The backtalk of self-generated sketches.
\newblock {\em {Design Issues}}, 19(1):72--88, 2003.

\bibitem{Gravetter12}
F.~Gravetter and L.-A. Forzano.
\newblock {\em Research methods for the behavioral sciences}.
\newblock Cengage Learning, 2012.

\bibitem{Heise14}
{Heise Verlag}.
\newblock Heise developer news.
\newblock \url{http://www.heise.de/developer/}.

\bibitem{Jamieson04}
S.~Jamieson.
\newblock {Likert scales: How to (ab)use them}.
\newblock {\em {Medical Education}}, 38(12):1217--1218, 2004.

\bibitem{LaToza06}
T.~D. LaToza, G.~Venolia, and R.~DeLine.
\newblock Maintaining mental models: a study of developer work habits.
\newblock In {\em {ICSE '06: 28th International Conference on Software
  Engineering}}, pages 492--501, 2006.

\bibitem{Lethbridge03}
T.~C. Lethbridge, J.~Singer, and A.~Forward.
\newblock How software engineers use documentation: the state of the practice.
\newblock {\em {IEEE Software}}, 20(6):35--39, 2003.

\bibitem{Mangano14}
N.~Mangano, T.~D. LaToza, M.~Petre, and A.~van~der Hoek.
\newblock Supporting informal design with interactive whiteboards.
\newblock In {\em {CHI'14: Proceedings of the SIGCHI Conference on Human
  Factors in Computing Systems}}, pages 331--340. ACM, 2014.

\bibitem{Myers08}
B.~Myers, S.~Y. Park, Y.~Nakano, G.~Mueller, and A.~Ko.
\newblock How designers design and program interactive behaviors.
\newblock In {\em {VL/HCC '08: Proceedings of the IEEE Symposium on Visual
  Languages and Human-Centric Computing}}, pages 177--184. IEEE, 2008.

\bibitem{Petre13}
M.~Petre.
\newblock {UML} in practice.
\newblock In {\em {ICSE '13: 35th International Conference on Software
  Engineering}}, pages 722--731. {IEEE}, 2013.

\bibitem{Petre14}
M.~Petre.
\newblock {\em Reflections on representations -- cognitive dimensons analysis
  of whiteboard design notations}, pages 267--294.
\newblock {Software Design in Action}. {CRC Press}, 2014.

\bibitem{Robbins11}
N.~B. Robbins and R.~M. Heiberger.
\newblock Plotting {L}ikert and other rating scales.
\newblock In {\em Proceedings of the 2011 Joint Statistical Meeting}, 2011.

\bibitem{Schuetze03}
M.~Sch\"utze, P.~Sachse, and A.~R\"omer.
\newblock Support value of sketching in the design process.
\newblock {\em {Research in Engineering Design}}, 14(2):89--97, 2003.

\bibitem{Searle00}
A.~Searle.
\newblock {\em Introducing research and data in psychology: A guide to methods
  and analysis}.
\newblock Routledge, 2000.

\bibitem{William02}
W.~R. Shadish, T.~D. Cook, and D.~T. Campbell.
\newblock {\em Experimental and quasi-experimental designs for generalized
  causal inference}.
\newblock Wadsworth Cengage Learning, 2002.

\bibitem{Spearman04}
C.~Spearman.
\newblock The proof and measurement of association between two things.
\newblock {\em {American Journal of Psychology}}, 15(1):72--101, 1904.

\bibitem{Suwa00}
M.~Suwa, J.~Gero, and T.~Purcell.
\newblock Unexpected discoveries and s-invention of design requirements:
  important vehicles for a design process.
\newblock {\em Design Studies}, 21(6):539--567, 2000.

\bibitem{Suwa02}
M.~Suwa and B.~Tversky.
\newblock External representations contribute to the dynamic construction of
  ideas.
\newblock In {\em {Diagrammatic Representation and Inference}}, pages 341--343.
  Springer, 2002.

\bibitem{Taylor07}
R.~N. Taylor and A.~Van~der Hoek.
\newblock Software design and architecture -- the once and future focus of
  software engineering.
\newblock In {\em {FOSE '07: International Conference on Software Engineering
  -- Workshop on the Future of Software Engineering}}, pages 226--243. IEEE,
  2007.

\bibitem{Tversky01}
B.~Tversky.
\newblock Spatial schemas in depictions.
\newblock In {\em {Spatial Schemas and Abstract Thought}}, pages 79--111, 2001.

\bibitem{Tversky02}
B.~Tversky.
\newblock What do sketches say about thinking?
\newblock In {\em {AAAI Spring Symposium, Sketch Understanding Workshop}},
  pages 148--151. Stanford University, 2002.

\bibitem{Tversky03}
B.~Tversky, M.~Suwa, M.~Agrawala, J.~Heiser, C.~Stolte, P.~Hanrahan, D.~Phan,
  J.~Klingner, M.-P. Daniel, P.~Lee, et~al.
\newblock {Sketches For Design and Design Of Sketches}.
\newblock In {\em {Human Behavior in Design: Individuals, Teams, Tools}}.
  Springer, 2003.

\bibitem{Ullman90}
D.~G. Ullman, S.~Wood, and D.~Craig.
\newblock The importance of drawing in the mechanical design process.
\newblock {\em {Computers \& Graphics}}, 14(2):263--274, 1990.

\bibitem{Hoek14}
A.~van~der Hoek and M.~Petre, editors.
\newblock {\em Software designers in action}.
\newblock CRC Press, 2014.

\bibitem{Walny11-2}
J.~Walny, S.~Carpendale, N.~Henry~Riche, G.~Venolia, and P.~Fawcett.
\newblock Visual thinking in action: visualizations as used on whiteboards.
\newblock {\em {Transactions on Visualization and Computer Graphics}},
  17(12):2508--2517, 2011.

\bibitem{Walny11-1}
J.~Walny, J.~Haber, M.~Dork, J.~Sillito, and S.~Carpendale.
\newblock Follow that sketch: lifecycles of diagrams and sketches in software
  development.
\newblock In {\em {VISSOFT '11: Proceedings of the 6th IEEE International
  Workshop on Visualizing Software for Understanding and Analysis}}, pages
  1--8. {IEEE}, 2011.

\bibitem{Wilcoxon45}
F.~Wilcoxon.
\newblock Individual comparisons by ranking methods.
\newblock {\em Biometrics}, 1(6):80--83, 1945.

\bibitem{Yatani09}
K.~Yatani, E.~Chung, C.~Jensen, and K.~N. Truong.
\newblock {Understanding how and why open source contributors use diagrams in
  the development of Ubuntu}.
\newblock In {\em {CHI '09: Proceedings of the SIGCHI Conference on Human
  Factors in Computing Systems}}, pages 995--1004. ACM, 2009.

\end{thebibliography}
\end{document}